\documentclass[prd,nofootinbib]{revtex4}
\usepackage{graphicx}
\usepackage{bm}
\def\be{\begin{equation}}
\def\ee{\end{equation}}
\def\ba{\begin{eqnarray}}
\def\ea{\end{eqnarray}}

\begin{document}

\title{\large \bf On vector mode contribution to CMB temperature and polarization 
from local strings}
\author{Levon Pogosian$^1$}
\author{Ira Wasserman$^{2,3}$}
\author{Mark Wyman$^{2,3}$}
\affiliation{
$^1$ Department of Physics, Syracuse University, Syracuse, NY 13244, USA \\
$^2$ Laboratory for Elementary Particle Physics,
Cornell University,
Ithaca, NY 14853, USA \\
$^3$ Center for Radiophysics and Space Research, Cornell
University, Ithaca, NY 14853, USA}
\date{\today}

\begin{abstract}
In a recent publication, we used the data from WMAP and SDSS to constrain the primordial
perturbations and to predict the B-mode polarization sourced by cosmic string networks.
We have been alerted \cite{Slosar} to the existence of errors 
in the code \cite{Pogosian} we used to calculate
the Cosmic Microwave Background anisotropies from cosmic strings. 
Correcting the errors leads
to a significant increase in the vector mode contribution
 to the CMB temperature and polarization anisotropies as well as an 
 overall renormalization of the various string spectra.
In these notes we explain the nature of the errors and discuss their implications 
for previously published constraints on cosmic strings based on this code. The chief change
in our results is that our derived limit for the cosmic string tension is
strengthened: $G\mu < 1.8 (2.7) \times 10^{-7}$ at 68 (95)\%
confidence. We also note that the newly-enhanced
vector mode contribution produces a greatly-increased amplitude for B-mode polarization
in the CMB which could exceed the B-mode power produced by the lensing of 
primordial E-mode polarization into B-mode polarization.
\end{abstract}


\maketitle

\section{Introduction}
Our collaboration has used the data from the Cosmic Microwave Background (CMB) anisotropy
and from galaxy surveys to place cosmological limits
on the properties of networks of cosmic strings in a series of papers \cite{PTWW03,PWW04,WPW05}.
We have recently learned \cite{Slosar} that the code we
used to calculate the cosmic string-sourced CMB anisotropy and
primordial power spectra contained several errors,
the most important of which was a missing normalization factor in the evaluation of 
the vector mode power produced by cosmic strings. This latter error alone lead 
to a factor of $\sim 8$ enhancement in vector-mode power. While revising the vector
mode part of the code we also located an overall factor of 2 normalization error in 
all string spectra. Full details of all corrected
errors are given in the Appendix.
 
The enhancement of the vector modes has implications for the results of our 
last paper \cite{WPW05}, especially for the prediction for the B-mode polarization
sourced by strings. The principal result of that work was a constraint on the
fractional power that strings can contribute to the CMB temperature anisotropy spectrum.
Fortunately, this fraction is mainly constrained 
by the shape of the string induced CMB spectrum, which did not change considerably 
as a result of fixing the code. Thus we believe that the bounds derived
in our last paper on the parameter we called $f$, the fractional 
power sourced by cosmic strings, will remain unchanged, and can 
still be used to derive bounds on the properties of the cosmic string network.
Changing the normalization of our curves, after correcting our errors,
scales downward our fiducial string tension $\mu_0$ by a factor of $1.8$,
so that our new fiducial string tension is $G\mu_0 \approx 1.1\times10^{-6}$.
The provisional limits on $G\mu$ that we derive using this corrected fiducial string tension are
then $G\mu < 1.8 (2.7) \times 10^{-7}$ at 68 (95)\% confidence.
The limits we placed on cosmic string substructure, parameterized by the string wiggliness, 
$\alpha_r$, were weak at best \cite{WPW05}; that determination should now be disregarded completely.
We plan to perform our Markov Chain Monte Carlo analysis again, with the 
corrected string spectra, using the new three-year WMAP data \cite{WMAP3}.
Until we finish this analysis, we will use these provisional bounds, which we 
expect to be at least approximately valid

A simple summary of what has changed in light of correcting these errors is that 
we now find much greater parity between the power in the vector and scalar modes 
caused by cosmic strings. Because this reapportionment of perturbation does not 
significantly alter
the shape of the string sourced CMB spectrum and, hence, should not affect the
cosmological constraints on the fraction of the total CMB power sourced
by cosmic strings, the greater proportion of vector mode power 
leads to a greatly enhanced prediction for B-mode polarization in the CMB.
Even with a smaller string tension, the B-mode power we now
predict is a factor of 10 - 20 times as great as reported in our previous work.
The factor of two uncertainty reflects our ignorance of cosmic string substructure. 
Given this enhancement, the B-mode polarization
signal for a cosmologically viable network of strings could be the most
prominent source for a B-mode signal in the CMB, with an amplitude even greater
than is expected from E-to-B lensing. A host of experiments aimed at measuring
B-modes with the relevant sensitivity ($\gtrsim 0.01 (\mu K)^2$ for $100 < \ell < 2000$)
are currently in planning or underway \cite{polarization}. The prominence of
the cosmic string B-mode prediction implies that these experiments
provide perhaps the best opportunity for either observing a network of cosmic strings
or placing limits on its properties.

\section{Cosmic Microwave Background Temperature Anisotropy}

\begin{figure}
\includegraphics[width=80mm]{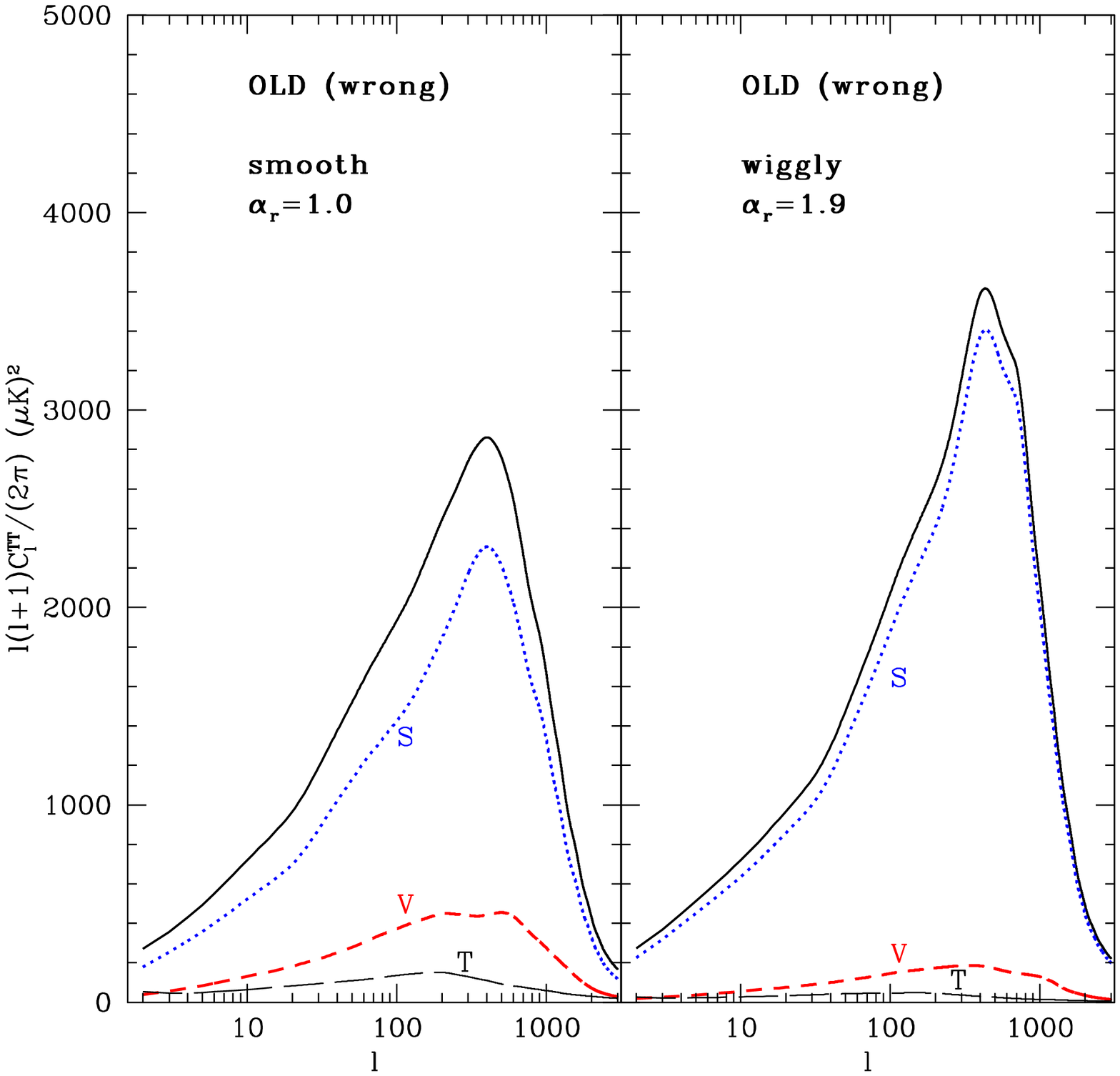}
\includegraphics[width=80mm]{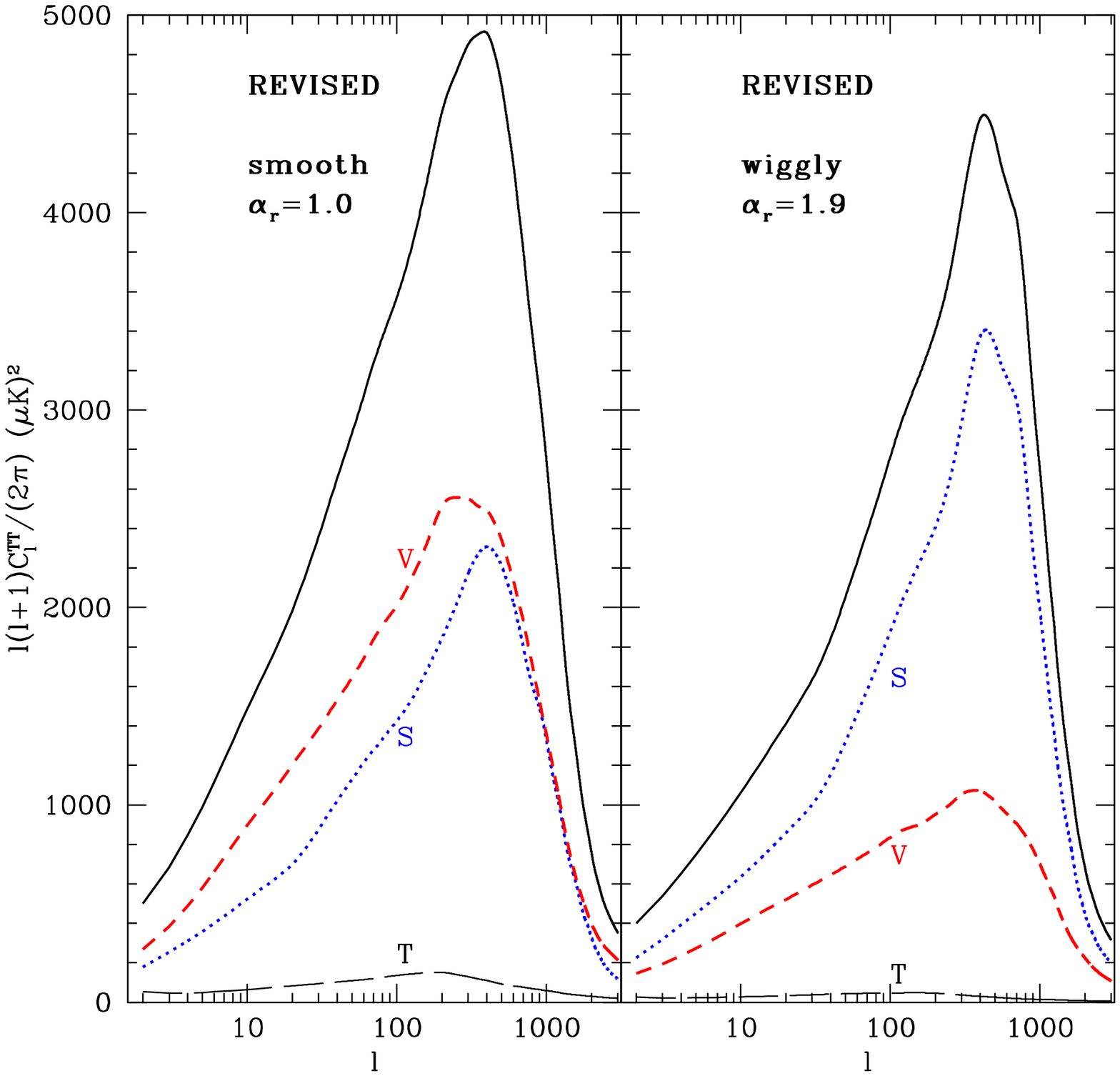}
\caption{\label{TT_svt} The scalar (blue dot), vector (red short dash), tensor (black long dash)
and their sum (solid black) contributions to the CMB TT spectrum evaluated using the old (left) 
and the revised (right) codes for the cases of smooth ($\alpha_r=1$) and wiggly ($\alpha_r=1.9$) 
strings with $G\mu = 1.4 \times 10^{-6}$. The spectra are proportional to $(G\mu)^2$ and 
this value of $G\mu$ corresponds to the reference cosmic string tension used in our previous work \cite{WPW05}, divided by $\sqrt{2}$ 
(to correct an overall factor of two missing in the power spectra computed from the 
previous version of our code).
Comparing the old and revised plots, it is evident that the new smooth string spectra have
a factor of about $1.69$ higher amplitude than the old spectra for the same string tension. 
We combine these two corrections -- a factor of $\sqrt{2}$ in overall normalization and a factor 
of $\approx \sqrt{1.69}$ from enhanced vector modes -- to infer that the correct, fiducial cosmic
string tension should be $G\mu_0 \approx 1.1 \times 10^{-6}$, which is a factor of $1.8$ smaller
than the fiducial tension estimated in our previous work. The meaning of this fiducial tension,
explained in our last paper, is this: $G\mu_0$ is the cosmic string tension that would be 
necessary for a network of cosmic strings to produce the same integrated primordial 
power as is found in the WMAP best-fit adiabatic model.
The parameter $\alpha_r$ is the ratio of the effective mass-per-length of a wiggly string
to that of a smooth string in the radiation era. See Refs. \cite{PV99,WPW05} for further details on string substructure.}
\end{figure}

As discussed in the Appendix, there was an overall factor of $2$ missing in
all spectra. One can correct for this factor by dividing the value
of the string tension $G\mu_0$ by a factor of $\sqrt{2}$. In particular, the
value $G\mu_0=2\times 10^{-6}$ used in our previous work \cite{WPW05}
 should now be replaced by  $G\mu_0=1.4 \times 10^{-6}$. 
In addition to this factor of $2$, the enhanced vector-mode power from 
our corrected code leads to a larger amplitude in the TT spectrum produced by cosmic strings
for a given cosmic string tension.
 In Fig.~\ref{TT_svt} we show the scalar, vector and tensor contributions
  to the string-sourced CMB TT spectrum, as well as their sum, before and
after correction. We now find that vector modes play a much more prominent role. In the
case of smooth strings their contribution is larger than that of scalar modes. This causes the total
TT power to be nearly doubled for the same value of $G\mu$. Wiggliness suppresses the vector modes
and enhances the scalar modes, a trend already pointed out in Ref. \cite{PV99}. The combined effect of wiggliness is to decrease the total TT power as compared with smooth strings. This means 
that constraints on $G\mu$ are {\emph{weaker}} for wiggly strings  than for smooth strings.
In contrast, our previous work \cite{PV99,WPW05} reported the erroneous conclusion that
wiggliness on strings produce an enhancement of total power and, therefore, a stronger
bound on $G\mu$. Thus, the conclusions reported in those papers need to be modified.
In particular, the bounds on string wiggliness reported there should be disregarded

It has been accepted among the experts in the field that global strings lead
to CMB anisotropies dominated by vector modes while local strings do not.
This presumption was, to a degree, based upon the results of previous studies
\cite{ABR99,PV99}, where vector modes were found to be subdominant even for
smooth strings. It now appears that this perception was incorrect, at least in the case of smooth strings.
However, we do expect local strings to be quite wiggly, 
since they accumulate small-scale structure over the course of their evolution. Hence,
we should still expect vector modes  to be relatively low for local strings.
Global strings, in contrast, are typically smooth and relativistic, so there is no
similar mechanism available to suppress their vector modes.

\section{B-mode Polarization in the CMB}

\begin{figure}
\includegraphics[width=80mm]{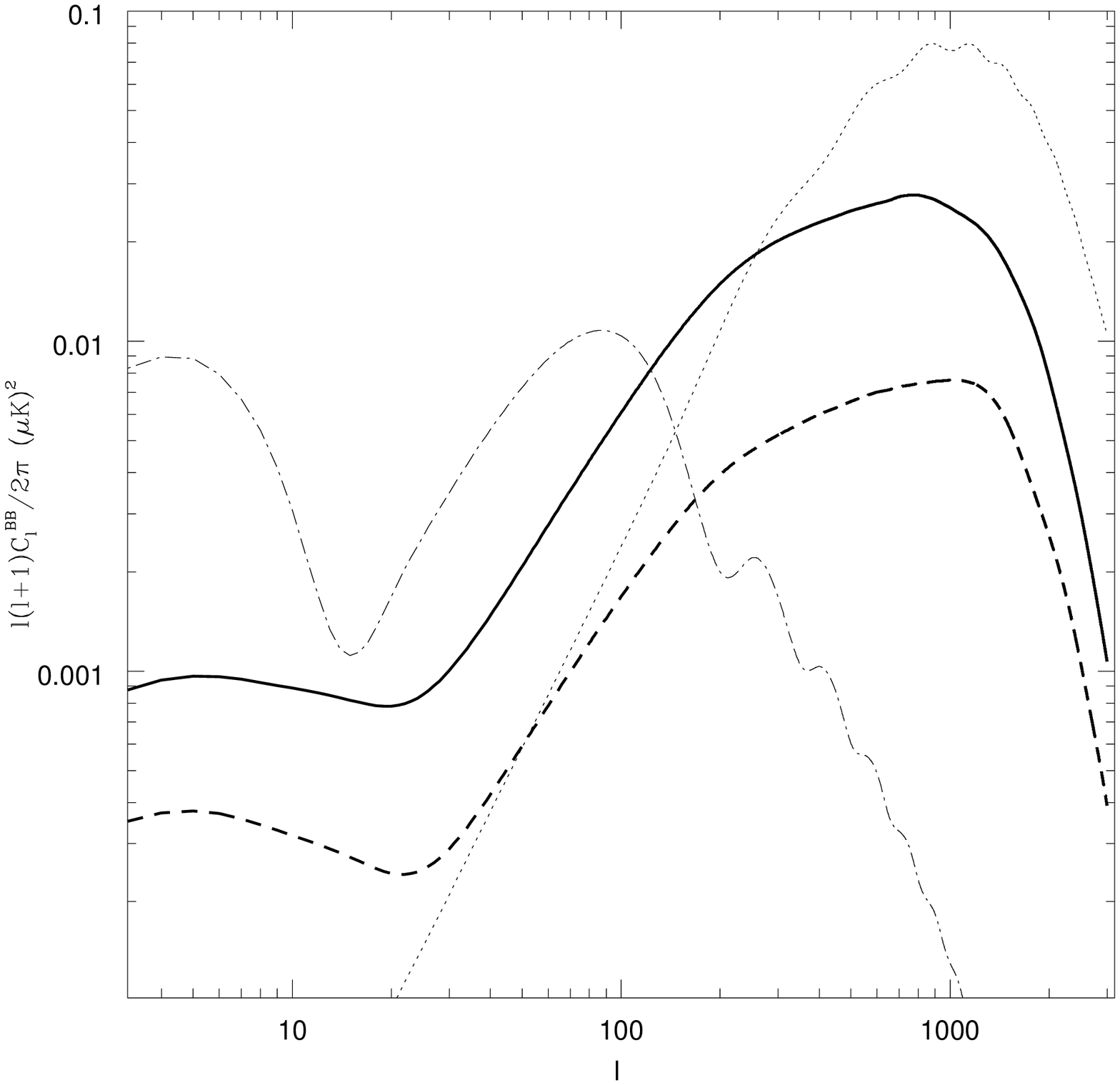}
\includegraphics[width=80mm]{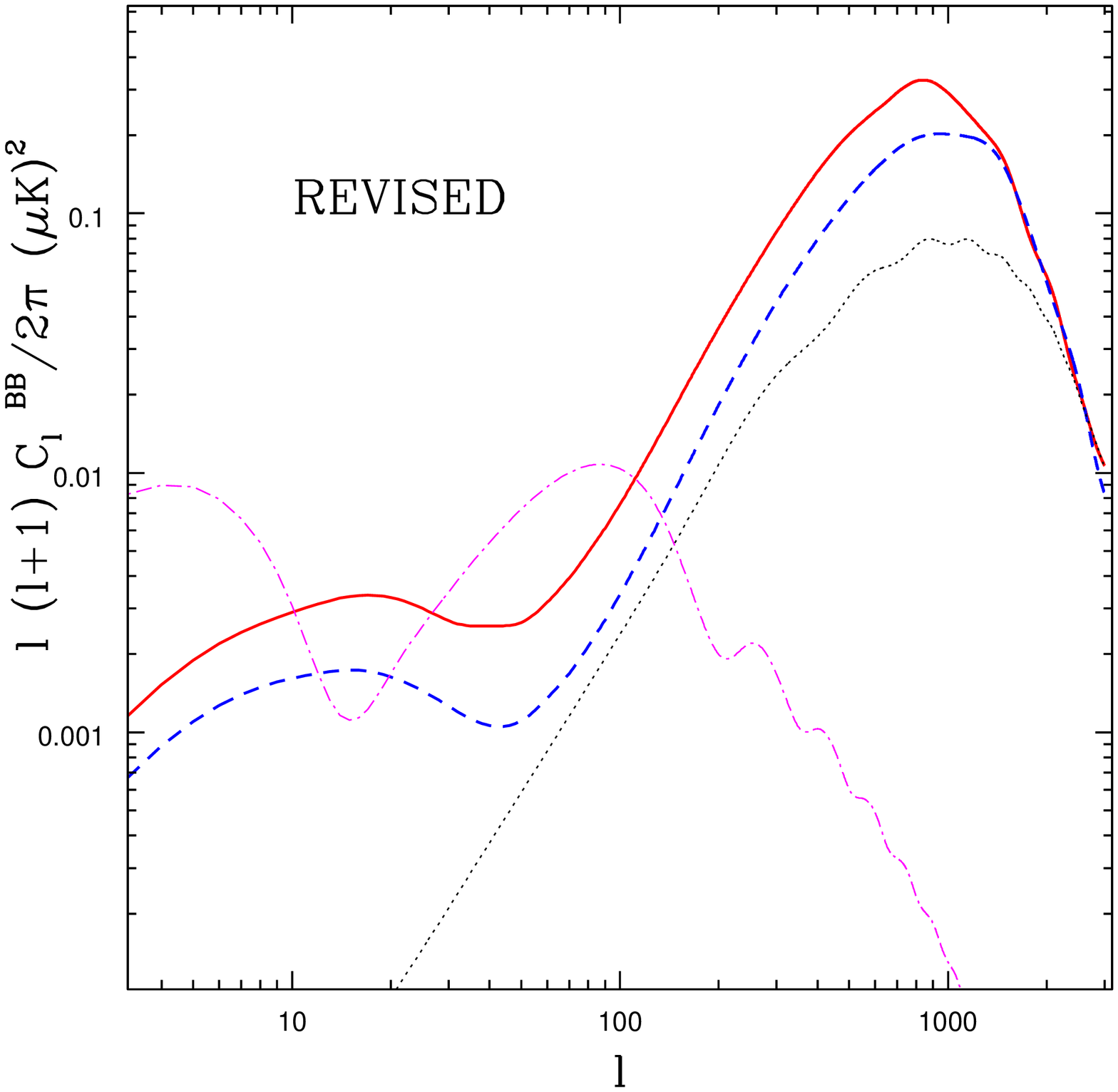}
\caption{\label{Cl_BB} The CMB BB spectra. Left: the plot from our last paper \cite{WPW05} showing
contributions from inflation for $r=0.1$ (dash-dot), lensing (dot) and cosmic strings (solid line for smooth and dash line for wiggly). Right: revised predictions for the smooth (solid, red)
and wiggly (dashed, blue) cases based on the assumption that strings contribute $10$\% of the power
in the CMB TT spectrum. Inflation and lensing estimates are plotted as in the left hand panel but with
the inflationary tensor mode color-coded magenta and E to B lensing still black.}
\end{figure}

Cosmic string-sourced vector mode perturbations can lead to
B-mode polarization in the CMB.
In view of the ongoing quest to observe B-type polarization in the CMB \cite{polarization},
it is of particular interest to revisit the B-mode predictions for cosmic strings
made in our previous analysis \cite{WPW05}.
In that paper we found that approximately $10$\% of the observed
CMB TT spectrum can be produced by strings (fraction in strings $f_s=0.1$). 
This fraction depends only on the shape of the string induced spectrum, so since
the shapes of the old and revised spectra, as shown in Fig.~\ref{TT_svt}, are quite similar,
we will assume that $f_s=0.1$ is still a valid upper bound.

Let us define $B$ as the coefficient by which one needs to multiply the
string TT spectrum in Fig.~\ref{TT_svt} in order for it to contribute $f_s=0.1$. 
The two spectra shown in Fig.~\ref{TT_svt} would need different values of $B$
to satisfy this bound. In our previous work \cite{WPW05} we determined that $f_s=0.1$ 
requires $B=0.1$ in the case of (uncorrected) smooth strings (as on the left panel of 
Fig.~\ref{TT_svt}). We can use this information to approximate, based on the ratios 
between the corrected and uncorrected $C_\ell$s, 
what values of $B$ are required for the revised spectra. For the corrected 
smooth string spectrum we estimate $B_{\rm sm}=0.06$, while for the corrected wiggly case 
$B_{\rm w}=0.07$.

Using these corrected $B$ factors, we are able to produce
revised estimates for the BB spectra. These are shown in Fig.~\ref{Cl_BB} along
with the plot from \cite{WPW05}. For a fixed $G\mu$, the net effect of correcting 
for the errors in the code was to boost the B-mode polarization from strings
by a factor of approximately $20$. Since the bounds on $G\mu$ have also been tightened 
as a result of our corrections, the effective increase in the
expected B-mode power is roughly a factor of $10$ in the smooth string ($\alpha_r =  1.0$) case. 
Because of compensating corrections, however,
the enhancement factor remains approximately $20$  in the wiggly string ($\alpha_r = 1.9$) case. 
Both predictions appear to be well above the estimated
limitations imposed by E to B lensing. A recent detailed study of the prospects of detection of the B-mode signal from cosmic strings, using a modification of our code, can be found in 
\cite{SS06}.

\section{Conclusion}
We have corrected several coding and normalization errors in the code we use to  calculate the CMB
anisotropy sourced by a network of cosmic strings. 
The principal changes caused by these corrections are an order of magnitude enhancement 
in the vector mode of primordial power spectra and an overall normalization
correction for the string spectra. Tracing through the implications of these errors,
we amend the principle results of our previous paper \cite{WPW05} and report a
more stringent cosmological bound on the cosmic string tension:  
$G\mu < 1.8 (2.7) \times 10^{-7}$ at 68 (95)\% confidence.
Besides tightening this bound, the increase in the vector power leads to another
implication. Cosmic strings can produce B-mode polarization in the CMB much more efficiently
than we estimated in our earlier work \cite{WPW05}
-- our previous predictions were too low by an order of magnitude.
The upshot of this correction is that we now predict that a
 cosmic string network that has not been excluded by current experiments can produce
a B-mode signal greater than any other anticipated source, including the lensing of inflationary 
E-mode polarization into B-mode polarization. This allows us to raise a tentative hope that
a cosmic string network, if it exists, could be observed through its B-mode polarization signal.
Alternatively, the non-observation of a cosmic string-sourced B-mode signal 
could provide the strictest observational limitation on the properties of such a network.

\acknowledgments

We thank Anze Slosar for alerting us to the problems with the code and useful discussions. We also acknowledge helpful discussions with Henry Tye and Tanmay Vachaspati.

\appendix
\section{The Errors}

\subsection{A missing factor of $2$ in the normalization of all CMB spectra}

This factor comes from the fact that the Fourier transform coefficients of the
string energy-momentum tensor are complex numbers, while the code
was evaluating only their real part. Since real and imaginary parts are
statistically equivalent, this was OK, as along as one used the real part
with the appropriate $\sqrt{2}$ correction to the amplitude. This factor
was inadvertently omitted in previous versions of our code.

\subsection{A missing factor of $-2\sqrt{2}$ in front of the vector source}
\label{2sq2}

The vector part of the cosmic string CMB code \cite{pogosian}
follows the conventions outlined in Ref. \cite{HW97}. The
conventions in the scalar and the tensor part of the code,
written as part of CMBFAST by Seljak and Zaldarriaga, are different and are described 
in Ref. \cite{ZS97}. In particular, the two conventions treat differently
the derivation of the numerical coefficients in front of the active sources on the RHS of 
the Einstein equations.

As implemented in the code, the cosmic string contribution to the vector modes comes 
through the $T_{13}$ component of the Fourier transform of the energy-momentum tensor
of the string network. This follows after one assumes ${\mathbf k}=\hat{\mathbf z}k$,
and the full argument for why there is no loss of generality in this procedure
can be found in Ref. \cite{ABR99}. Here we derive the relation between 
$T_{13}({\mathbf k},\eta)$ (which corresponds to the variable $\rm emtV$ in the code)
and the vector anisotropic stress $\pi^{(1)}_s({\mathbf k},\eta)$ that appears on the RHS 
of Eq.~(70) of \cite{HW97}.

In Ref. \cite{HW97}, $\pi^{(1)}_s({\mathbf k},\eta)$ is defined as the coefficient of the
expansion of $T_{ij}({\mathbf x},\eta)$ in the set of eigenfunctions 
$Q^{(1)}_{ij}({\mathbf k},{\mathbf x})$:
\be
Q^{(1)}_{ij}({\mathbf k},{\mathbf x})=-{1 \over 2\sqrt{2}}
\left[ \hat{k}_i (\hat{e}_x + i\hat{e}_y)_j + \hat{k}_j (\hat{e}_x + i \hat{e}_y)_i \right] 
e^{i{\mathbf k}{\mathbf x}} \ .
\label{qij}
\ee
The relevant equation in Ref. \cite{HW97} is Eq.~(40), which says
\be
T_{ij}= \pi^{(1)}_s({\mathbf k},\eta) Q^{(1)}_{ij}({\mathbf k},{\mathbf x}) \ .
\label{tij}
\ee
This is perhaps a bit confusing. One should, in principle, write the full expansion:
\be
T_{ij}({\mathbf x},\eta)=\sum_{\mathbf k}
\left[ (Q^{(0)}\delta_{ij}+Q^{(0)}_{ij})p_s +
\sum_{m=\pm 1} \pi^{(m)}_s Q^{(m)}_{ij} + \sum_{m=\pm 2} \pi^{(m)}_s Q^{(m)}_{ij}
\right]
\ee
The vector part of the energy-momentum tensor $T_{ij}$ has contributions from 
both $\pi^{(1)}_s$ and $\pi^{(-1)}_s$. However, the function $V({\mathbf k},\eta)$
on the LHS of Eq.~(70) of Ref. \cite{HW97} was also defined in Eq.~(37) only in
terms of $Q^{(1)_i}$. As the authors say, this is acceptable, since $+$ and $-$ modes are
equivalent and it is sufficient to consider just one of them, while taking proper
care of numerical prefactors. In the code, Eq.~(37) of Ref. \cite{HW97} was interpreted as
\be
h_{0j}({\mathbf k},\eta) e^{i{\mathbf k}{\mathbf x}} = - V Q^{(1)}_{j} \ ,
\ee
where $h_{0j}({\mathbf k},\eta)$ is the Fourier transform of $h_{0j}({\mathbf x},\eta)$.
As a result, one obtains the following equation for $V({\mathbf k},\eta)$:
\be
Q^{(1)}_{ij} k \left[ \dot{V} + 2 {\dot{a} \over a} V \right] = 
- 8 \pi G a^2 \left[ p_f \pi^{(1)}_f Q^{(1)}_{ij} +T_{ij} e^{i{\mathbf k}{\mathbf x}} \right] \ . 
\label{deriveV}
\ee
Choosing ${\mathbf k}=\hat{\mathbf z}k$ assures that $T_{13}$ and $T_{23}$ are pure vector
modes ({\it i.~e.} they do not contribute to the scalar and tensor modes).
For the $\{13\}$ component, using Eq.~(\ref{qij}) with ${\mathbf k}=\hat{\mathbf z}k$, we have
\be
T_{13} e^{i{\mathbf k}{\mathbf x}} = 
- T_{13} {2 \sqrt{2} \over (\hat{e}_x + i\hat{e}_y)_1} Q^{(1)}_{13} = -2 \sqrt{2} T_{13} Q^{(1)}_{13}   \ ,
\label{tijx}
\ee
where in the last step we used the fact that 
$(\hat{e}_x+ i\hat{e}_y)_1 \equiv (\hat{e}_x + i\hat{e}_y)_x = 1$. Hence, using (\ref{tijx})  
in (\ref{deriveV}), we can write
\be
k \left[ \dot{V} + 2 {\dot{a} \over a} V \right] = 
- 8 \pi G a^2 \left[ p_f \pi^{(1)}_f - 2\sqrt{2} T_{13} \right]  \ . 
\label{finalV}
\ee
It was the factor of $-2\sqrt{2}$ in front of $T_{13}$ that was missing in the code,
as was pointed out to us by Slosar \cite{Slosar}. In other
words
\be
\pi^{(1)}_s({\mathbf k},\eta) = -2\sqrt{2} T_{13}({\mathbf k},\eta) \ .
\ee
Note that this would lead to a factor of $8$ increase in the vector mode contribution to
the TT and BB power spectra. 

This factor would not be necessary if one followed the conventions of scalar-vector-tensor (SVT) decomposition used in Ref. \cite{TPS97}. That same decomposition was used in Ref. \cite{ABR99}. This is probably the reason for the aforementioned oversight, since a cross-check against results of Ref. \cite{ABR99} was one of our tests. It now appears that the vector modes 
in Ref. \cite{ABR99} were also underestimated, possibly through a similar error in normalization.
The tensor part of CMBFAST uses the SVT decomposition consistent with Ref. \cite{TPS97} and
does not require additional factors in front of $T_{ij}$.

\subsection{Other Coding Errors}

Some other errors were discovered in the vector part of the
code. One was essentially a typo, whose effect was to set the time derivative of
the scale factor during tight coupling to zero. More specifically, in {\tt subroutine fderivsv},
instead of 
\be
{\tt do} \ {\tt 80} \ {\tt l=1,nv} 
\ee
it should have been 
\be
{\tt do} \ {\tt 80} \ {\tt l=3,nv} \ .
\ee 
The effect of fixing this error was a {\emph decrease} in the polarization sourced by vector
modes.

The other errors were in
{\tt subroutine foutputv}, in the equations that evaluates variables {\tt dve} and {\tt dvb},
which are the E- and B-mode source functions related to variables $S^{(1)}_1$ and $S^{(1)}_2$ defined in
Eq~(61) of Ref. \cite{HW97}.
The old equations were
\ba
\nonumber
{\tt dve} &=& {\tt (-sqrt(6.0d0)/10.0d0)*( vis(j)*polv/x**2
+ (vis(j)*polvpr + dvis(j)*polv)/(ak*x)  ) }
\\
{\tt dvb} &=& {\tt (-sqrt(6.0d0)/10.0d0)*vis(j)*polv/x}
\ea
The corrected equations are
\ba
\nonumber
{\tt dve} &=& {\tt (-sqrt(6.0d0)/2.0d0)*(2.0d0*vis(j)*polv/x**2
+ (vis(j)*polvpr + dvis(j)*polv)/(ak*x)  ) }
\\
{\tt dvb} &=& {\tt (-sqrt(6.0d0)/2.0d0)*vis(j)*polv/x}
\ea
Namely, there was an extra prefactor of $1/5$, which was inserted thanks to
a confusion caused by a difference in the conventions used by Refs. \cite{HW97}
and \cite{ZS97}, and a missing factor of $2$ in front of the
first term in the equation for {\tt dve}, which was a result of a careless
integration by parts.

The combined effect of fixing the errors specified in this subsection was
to increase the amplitude of both the E and B polarization sourced by the
vector modes by a factor of about 2-3 in their power spectra.

\end{document}